# Phonons and Hybrid Modes in the High and Low Temperature Far Infrared Dynamics of Hexagonal $TmMnO_3$

Néstor E. Massa*.[1] Leire del Campo,[2] Domingos De Sousa Meneses,[2] Patrick Echegut,[2] María Jesús Martínez-Lope,[3] and José Antonio Alonso[3]

[1] Laboratorio Nacional de Investigación y Servicios en Espectroscopía Óptica-Centro CEQUINOR, Universidad Nacional de La Plata, C.C. 962, 1900 La Plata, Argentina.

[2]CNRS, CEMHTI UPR3079, Université d'Orléans, F-45071 Orléans, France.

[3]Instituto de Ciencia de Materiales de Madrid, CSIC, Cantoblanco, E-28049 Madrid, Spain.

.

•e-mail: neemmassa@gmail.com

# ABSTRACT

We report on temperature dependent $TmMnO_3$ far infrared emissivity and reflectivity spectra from 1910 K to 4 K. At the highest temperature the number of infrared bands is lower than that predicted for centrosymmetric $P6_3/mmc$ ($D_{6h}^4$) (Z=2) space group due high temperature anharmonicity and possible defect induced bitetrahedra misalignments. On cooling, at ~1600 ± 40 K, $TmMnO_3$ goes from non-polar to an antiferroelectric-ferroelectric polar phase reaching the ferroelectric onset at the ~700 K.

Room temperature reflectivity is fitted using 19 oscillators and this number of phonons is maintained down to 4 K. A weak phonon anomaly in the band profile at 217 $cm^{-1}$ (4 K) suggests subtle Rare Earth magnetoelectric couplings at ~$T_N$ and below.

A low energy collective excitation is identified as a THz instability associated with room temperature $e_g$ electrons in a d-orbital fluctuating environment. It condenses into two modes that emerge pinned to the E-type antiferromagnetic order hardening simultaneously down to 4 K. They obey power laws with $T_N$ as the critical temperature and match known zone center magnons. The one peaking at 26 $cm^{-1}$, with critical exponent β=0.42 as for antiferromagnetic order in a hexagonal lattice, is dependent on the Rare Earth ion. The higher frequency companion at ~50 $cm^{-1}$, with β=0.25, splits at ~$T_N$ into two peaks. The weaker band of the two is assimilated to the upper branch of gap opening in the transverse acoustical (TA) phonon branch crossing the magnetic dispersion found in $YMnO_3$. (Petit et al, 2007 Phys. Rev. Lett. **99**, 266604). The stronger second at ~36 $cm^{-1}$ corresponds to the lower branch of the TA gap. We assign both excitations as zone center magnetoelectric hybrid quasiparticles concluding that in $NdMnO_3$ perovskite the equivalent picture corresponds to an instability which may be driven by an external field to transform $NdMnO_3$ into a multiferroic compound by perturbation enhancing the TA phonon-magnetic correlation.

**PACS #**

Phonons 74.25.Kc

Phase transition ferroelectric, 77.80.B-

Collective effects 71.45.Gm

Infrared spectra 78.30.Hv

Emissivity 78.20.Ci (far infrared)

Magnetoelectric effects, multiferroics 75.85.+t

Phonon interactions with other quasiparticles 63.20.kk

# INTRODUCTION

From the initial days of research in oxides [1,2] it was recognized that hexagonal rare earth manganites conform an attractive group with a wide range potential practical applications.[3,4] One main distinctive feature of this family of compounds is the coexisting ferroelectric and magnetic order without the occurrence of the electronic lone pair found in classical ferroelectrics at the perovskite site A (Bi, Pb ions) in which two valence electrons participate in chemical bonds through (sp)-hybridized states such as $sp^2$ or $sp^3$ [5]

From a plethora of experimental probes on these perovskite derived improper ferroelectrics,[6] studies of the vibrational and electronic properties through the paraelectric, ferroelectric, and magnetic phases unveil clues on the weight and role of the intrinsic interplay of orbital, spin, lattice, and charge. Among them, hexagonal $RMnO_3$ (R=Rare Earth) undergo a paraelectric to ferroelectric transition by losing the inversion center of the high temperature space group $P6_3/mmc$. The spontaneous electric polarization along **_c_** axis is triggered by buckling in the **_ab_** plane accompanied by ferroelectric Rare Earth displacements to inequivalent sites.[7]

Sharing these properties, $TmMnO_3$¸ occupies an intermediate place within the family of compounds with smaller Rare Earth cations (R= Lu, Yb, Tm, Er, Ho).[8] Its hexagonal crystal structure is built on corner-sharing bipyramids in which center lies the $Mn^{3+}$ ion surrounded by three in-plane oxygens. The $MnO_5$ polyhedra, with the two extra apical $O^{2-}$ ions along the c axis, define a near triangular network in the **_ab_** plane where $Mn^{3+}$ long-range frustrated triangular antiferromagnetism, due to a ~120° spin arrangement, sets in at $T_N^{Mn}$~84 K.[9,10] In this lower temperature phase, $TmMnO_3$ has six formula units per unit cell. Neighboring spins are antiferromagnetic coupled via oxygen ions by superexchange interactions. And, as in any other $RMnO_3$ (R=Rare Earth), interplane exchange interactions are expected to be about two orders of magnitude weaker [11] but still play a significant role in the origin of multiferroicity.[12, 13]

Thulium ions are in cages surrounded by seven oxygens ($TmO_7$) in a trivalent state and occupy two non-equivalent lattice sites. It is anticipated to also order magnetically at an unspecified lower temperature $T_N^{Th.} \ll T_N^{Mn}$ [9, 14]

Here we present temperature dependent far- and mid- infrared emissivity and reflectivity of hexagonal $TmMnO_3$ from 1900 K to 4K exploring the onset of structural changes leading to polar and magnetic order and mode hybridization. Starting at about 1900 K, and similarly to $YMnO_3$,[15] the structural transition in $TmMnO_3$ from high temperature non-polar group $P6_3/mmc$ to the hexagonal space group $P6_3cm$, underlines the loss of the lattice inversion center on cooling toward the polar state. At the 800-600 K temperature range the inversion center loss is evident in the increase of intensity of the 230 $cm^{-1}$ O-Mn-O vibrational bending band. Changes at antisymmetric and symmetric internal vibrational frequencies denote further octahedral deformations as $TmMnO_3$ gradually becomes truly polar and ferroelectric.[16]

Cooling further down to 80 K, and concurrent with the observation of no structural changes, the number of infrared active phonons remains constant. Below that temperature most of the vibrational bands undergo temperature driven sharpening as expected for cooling.

In addition to phonon activity, we report on the temperature evolution of THz excitations represented by a room temperature overdamped band, the origin of which is assigned to orbital fluctuating $e_g$ electrons. Starting at about 100 K, and notoriously, at $T_N$~84 K, the band comes out steadily toward higher frequencies and allows modes as for a second order transition mirroring the development of the long-range spin ordering in the frustrated ***ab*** plane anisotropic hexagonal topology. At about the antiferromagnetic onset it locks into two main features with the same energy scale as known zone center magnons. The lower energy band at 26 $cm^{-1}$ is rare earth dependent while the stronger one centered at ~50 $cm^{-1}$ splits into two subbands that may be assimilated to the upper and lower branch of a transverse acoustical phonon gap inducing zone center hybridized acoustical phonon-magnetic dispersion quasiparticles.[17]

## SAMPLE PREPARATION AND EXPERIMENTAL DETAILS

$TmMnO_3$ was prepared as polycrystalline powder by a liquid-mix technique. Stoichiometric amounts of analytical-grade $Tm_2O_3$ and $MnCO_3$ were dissolved in citric acid by adding several droplets of concentrated $HNO_3$ to favor the solution of $Tm_2O_3$; the citrate + nitrate solution was slowly evaporated, leading to an organic resin that was first dried at 120 °C, and then, decomposed by heating at temperatures up to 800 °C in air. The precursor powder was finally heated at 1100 °C in air for 12 h, thus yielding a well-crystallized single phase powder. [11, 18].

The $TmMnO_3$ crystal structure at room temperature is defined in the hexagonal space group $P6_3cm$ with the lattice parameters a= 6.0950(1) and c= 11.3806(2) Å. [11, 18]. As it is shown in Fig. 1 (b) each Mn atom is coordinated by five oxygen atoms in a bipyramidal configuration. One O3 atom and two O4 atoms are in the equatorial plane of the bypyramid, whereas the O1 and O2 atoms are at the apexes. Tm atoms occupy two crystallographic positions, *Tm1* and *Tm2*, bonded each to seven oxygen atoms. Both $TmO_7$ polyhedra can be described as monocapped octahedra. The capping oxygen is O3 for Tm1 and O4 for Tm2. Along the z axis, the structure consists of layers of corner-sharing $MnO_5$ bypiramids separated by layers of edge-sharing $TmO_7$ polyhedra.

Temperature dependent medium (MIR), and far infrared (FIR) near normal reflectivities from 4 K to 850 K of polished high quality polycrystal samples in the shape of 10 mm diameter pellets were measured with a FT-IR Bruker 113v interferometer at 2 $cm^{-1}$ resolution. Supporting measurements between 2 $cm^{-1}$ and 80 $cm^{-1}$ have also been done in a Bruker 66v/S interferometer at the IRIS-Infrared beamline of the Berlin Electron Storage Ring (BESSYII-Helmholtz Zentrum Berlin für Materialien und Energie. GmbH). Sample and reference were placed in the sample chamber inside the spectrometer in equivalent positions as related to the placement of the detector and the ratio between sample and reference signals gave us the absolute reflectivity spectra.

For high temperature reflectivity (up to ~850 K) we used a heating plate adapted to the near normal reflectivity attachment in the Bruker 113v vacuum chamber. In this temperature range, the spurious infrared signal introduced by the hot sample thermal radiation was corrected to obtain the reflectivity values. All measurements were taken on heating runs.

For the phonon normal emissivity, the ratio of the sample luminescence ($\mathscr{L}_S$) relative to the black body's ($\mathscr{L}_{BB}$), was measured with two Fourier transform infrared spectrometers, Bruker Vertex 80v and Bruker Vertex 70, optically coupled to a rotating table placed inside a dry air box. This allows measuring spectral emittance in two dissimilar ranges from 20 $cm^{-1}$ to 7000 $cm^{-1}$ simultaneously. The sample was heated with a 500 W pulse Coherent $CO_2$ laser.[19,20,21]

Measured emissivity spectra are obtained by applying the following expression.[22]

$$\boldsymbol{E}(\omega,T)=\frac{FT\left(\mathrm{I_S}-\mathrm{I_{RT}}\right)}{FT\left(\mathrm{I}_{BB}-\mathrm{I}_{RT}\right)}\times\frac{\mathscr{P}\left(\mathrm{T}_{BB}\right)-\mathscr{P}\left(\mathrm{T}_{RT}\right)}{\mathscr{P}\left(\mathrm{T}_{S}\right)-\mathscr{P}\left(\mathrm{T}_{RT}\right)}\boldsymbol{E}_{BB} \quad (1)$$

where *FT* stands for Fourier Transform., and I for measured interferograms i.e., sample, $I_S$ ; black body, $I_{BB}$ ; and, environment, $I_{RT}$ . $\mathscr{P}$ is the Planck's function taken at different temperatures T; i.e., sample, $T_S$; blackbody, $T_{BB}$; and surroundings, $T_{RT}$. $\boldsymbol{E}_{BB}$ is a correction that corresponds to the normal spectral emissivity of the black body reference (a $LaCrO_3$ Pyrox PY 8 commercial oven) and takes into account its non-ideality.

A liquid He cooled bolometer and a deuterated triglycine sulfate (DTGS) detector were employed to completely cover the spectral range of interest. A gold mirror was used as 100% reflectivity reference since emissivity and reflectivity spectra are in excellent agreement in our range of interest. This makes possible the use of the same sample with both techniques without altering its surface (by running first reflectivity in vacuum and then emissivity in dry air (see below)).

After obtaining the optical data we placed all our emission spectra in a more familiar near normal reflectivity framework by noting that

$$\mathrm{R}=1-\mathrm{E} \quad (2)$$

where $\mathrm{R}$ is the sample reflectivity. This allows computing phonon frequencies using a standard multioscillator dielectric simulation.[24] We use a description of the dielectric function, $\varepsilon(\omega)$, given by

$$\varepsilon(\omega)=\varepsilon_1(\omega)-i\varepsilon_2(\omega)=\varepsilon_\infty\prod_j\frac{(\omega_{jLO}^2-\omega^2+i\gamma_{jLO}\omega)}{(\omega_{JTO}^2-\omega^2+i\gamma_{jTO}\omega)} \quad (3)$$

$\varepsilon_\infty$ is the high frequency dielectric constant taking into account electronic contributions; $\omega_{jTO}$ and $\omega_{jLO}$, are the transverse and longitudinal optical mode frequencies and $\gamma_{jTO}$ and $\gamma_{jLO}$ their respective damping. It is also proper to mention that the left-hand side of equation (3) is, in fact, an approximation strictly valid for magnetically inert materials. Its use implies that the unknown frequency dependent magnetoelectric coupling constant, $\alpha(\omega)$, is set to zero. On the other hand, magnetoelectric effects are very weak and in our case corrections to the phonon Lyddane-Sachs-Teller relationship are expected to be negligible.[24] Our measurements suggest that that coupling constant may be magnon-phonon dependent.

We also added when needed a Drude term (plasma contribution) to the dielectric simulation as

$$-\frac{\left(\omega_{pl}^2 + i\cdot\left(\gamma_{pl}-\gamma_0\right)\cdot\omega\right)}{\left(\omega - i\gamma_0\right)\cdot\omega} \qquad (4)$$

where $\omega_{pl}$ is the plasma frequency, $\gamma_{pl}$ its damping, and $\gamma_0$ is understood as a phenomenological damping introduced to reflect lattice drag effects. When these two dampings are set equal, one retrieves the classical Drude formula.

The real ($\varepsilon_1(\omega)$) and imaginary ($\varepsilon_2(\omega)$) part of the dielectric function (complex permittivity, $\varepsilon^*(\omega)$) then estimated from fitting the data [25] using the reflectivity R given by

$$\mathrm{R}(\omega) = \left|\frac{\sqrt{\varepsilon^*}(\omega)-1}{\sqrt{\varepsilon^*}(\omega)+1}\right|^2 . \qquad (5)$$

We calculated the oscillator strength $S_j$ for the $j^{th}$ oscillator as in

$$S_j = \omega_{jTO}^{-2} \frac{\prod_k (\omega_{kLO}^2 - \omega_{jTO}^2)}{\prod_{k \neq j} (\omega_{kTO}^2 - \omega_{jTO}^2)} \qquad (6)$$

## Results and Discussion

### *i) The non-polar to polar high temperature structural phase transition*

Fig. 2 shows the 1-E spectra of $TmMnO_3$ from 662 K to 1910 K. We find that despite the factor analysis of the space group $P6_3/mmc$ ( $D_{6h}^4$) (Z=2) the high temperature centrosymmetric structure predicts [18]

$$\Gamma_{IR} = 3A_{2u} + 3\ E_{1u} \qquad (7)$$

infrared active modes, our spectra display fewer bands (Table I). Being weak shoulders or very broad, they may be interpreted as unresolved due to high anharmonicities in paraelectric $TmMnO_3$. It is however worth mentioning that as in some high temperature compounds with basal plane of hexagonal structure there might be up/down vertical sublattice orientational distortion contributing to band broadening.[26, 27] This would then add an intrinsic extra factor associated with possible bitetrahedra misalignments in passing from highest temperature centrosymmetric P63/mnm to the $P6_3cm$ space group.

On cooling, we find a clear discontinuity in the number of phonons and band profiles at 1600 K ± 40 K. (Fig. 2, inset) where new band and relative intensity changes are assimilated to conventional order-disorder mechanism for a first order phase transition. We assign that phonon change to the transition from non-polar to polar structure observed by Lukaszewicz and Karat-Kaliciníska, and others in $YMnO_3$ [28. 29] Further, our spectra and analysis at slightly higher temperature (our working temperatures are fixed by the $CO_2$ laser ramp), support the scenario proposed for $TmMnO_3$ by Lonkai et al using high temperature neutron powder-diffraction data.[15] Antiferroelectricity and canted ferroelectricity is proposed as part of a two-step structural transition toward the low temperature polar phase. Then at $T_{npl}$~1600 K vibrational changes, shown in Fig.. 3 (a,b) below 1000 $cm^{-1}$ for 1-E at 1760 K and 1382 K, relate to $MnO_5$ rigid bipyramid tilts corrugating the rare earth layer. I. e., in the antiferroelectric phase phonon changes are compatible with individual Rare Earth displacements in opposite directions compensating in balance the ferroelectric polarization.[15] Table I shows the number of oscillators needed to get an excellent fit on both sides of the transition.

At temperatures lower than Tc, the reflectivity at 580 K and 4 K in the $P6_3cm$ hexagonal phase (fig.3 c, d) show the result of unit cell tripling with six formula units as a clear increment of the number of better defined oscillators (Table I).

We also found in 1-E a remarkable change in a very broad band centered at frequencies higher than ~6000 $cm^{-1}$ in the mid-infrared spectra at about the temperatures where the phase transition takes place (Fig. 2) This band remains well defined below ~1600 K, and remarkably, is not detected in the centrosymmetric phase where one can only infer a shifted weight toward lower frequency. Although at these temperatures in the mid-infrared spectral region may be argued that eq.(2) is not strictly valid due to the fact that transmission is not taken into account, thus introducing the possibility of some profile distortion, it should be also kept in mind that being our samples oxides, they are subjected to atmospheric oxidation when heated in a dry air box. Discussed in detail for $NdMnO_3$ [20] heating in air is known to generate a mid-infrared band by carrier delocalization from the oxidation reaction $Mn^{3+} \rightarrow Mn^{4+} + 1e^-$, independently of the phase at which it takes place.[20]

As it is shown in Fig. 3 (a) and table I, the spectrum fit for 1-E in the centrosymmetric phase, above the ~1600 K phase transition, needs the explicit introduction of an overdamped Drude term indicating the presence of leakage currents by defects. The Christiansen inflexion point [20] that it is still present at about 750 $cm^{-1}$ now is at a finite reflectivity distorted by the appearance of somehow delocalized carriers in the background. Then its use in calculating the sample temperature (eq(1)) is not reliable as for temperature reading. [20]. As it is shown in fig. 2 this reflectivity dip becomes even much weaker after further temperature increase beyond ~1910 K where $TmMnO_3$ becomes a poor conducting compound as a consequence of oxidation defects inducing hopping of quasi-free electrons These electrons are associated with the double exchange mechanism between $Mn^{3+}$ and $Mn^{4+}$ ions. One might then be inclined to conclude that the intensity enhancement found in the ~6000 $cm^{-1}$ band abruptly changing at $T_{npl}$, is result of the lattice distortion (symmetry) and temperature driven carrier localization affecting the electronic polarization of the lattice ions. That is, electron-dipole cooperative interactions among defect induced small polarons, quasilocalized carriers, and bond polarization arise as the system moves below ~1600 K into polar order in the ferroelectric environment.

### *ii) The $P6_3cm$ ferroelectric distortion*

Fig. 4 shows that the onset of ferroelectricity in $TmMnO_3$ appears beginning at ~800 K due to additional bipyramid distortion [8] when, upon cooling the lattice, it gradually goes from triangular antiferroelectric to triangular ferroelectric within the $P6_3cm$ space group.[15**]** **T**he emergent polar environment implies better defined vibrational bands The most distinctive change takes place at ~300 $cm^{-1}$, which is the frequency for a "bending-antisymmetric" mode linking O-Mn distances in the polyhedron non-equivalent local deformation. It implies modifications in the O(1)-Mn-O(2) sublattice distances related to the spontaneous electric polarization. This mode is traceable to the primordial silent cubic infrared distortion activated by the unit cell tripling and lower symmetry enhancing the $MnO_5$ bipyramid tilting. This may also include contributions from Mn $t_{2}g$ orbitals.[30] Its new activity in the 800K-600 K range also agrees with the temperature of a dielectric anomaly at ~621 K reported by Wang et al for the ferroelectric distortion.**[**16**]**

### *iii) Lower temperature phonons*

Factor analysis of the irreducible representation of the space group $P6_3cm$ (Z=6) for the room temperature ferroelectric phase in $TmMnO_3$ yields

$$\Gamma_{IR} = 9A_1 + 14E_1 \qquad (8)$$

That is, being the lattice non-centrosymmetric, the 23 vibrational modes belonging to the $A_1$ and $E_1$ species are simultaneously infrared and Raman active.[18]

Fig. 3 (d) and table I show that only 19 oscillators are required for getting an excellent fit in this phase, probably, because of broadening and near degeneracy. No new optical phonon modes are observed on cooling through the antiferromagmetic transition at $T_N$ ~84 **K** from 300 K to 4 K **(**Fig 5).

However, the band for the lattice mode at ~219 $cm^{-1}$ and 4K (Fig. 6 and table 1).involving Rare Earth displacements in the $TmO_7$ cages adjacent to the bipyramids has a stronger temperature dependent profile (Fig. 6, upper panel) that is better modeled in the Néel phase adding a new oscillator as side band (~237 $cm^{-1}$ at 4 K, Table I). This aims to account for possible minute changes in the non-equivalent Tm lattice sites associated with the activation of the superexchange Tm-O-Mn pathways, off the ***ab*** plane, where tetrahedra distortions are known to be consequence of non-equivalent Rare Earth sites (Fig. 1(b) [14]. Then, the electric dipole change would signal coupling between the magnetic and electric counterparts [31] in an environment with shorter local phonon correlation length yielding a dispersion whose contribution, close but off-zone center, distorts the infrared vibrational profile. It is also interesting to note that this consequence, related to Rare Earths, may correlate to earlier optical studies for $YbMnO_3$ where it has been observed exchange line splitting symmetrically into two at 85 K in the $Yb^{3+}$ crystal field spectrum following its magnetization curve yielding $T_N$= 87.3K [32]

### *iv) THz hybrid soft modes*

Overall, magnetoelectric coupling in hexagonal $TmMnO_3$ is expected to result in a more relaxed lattice setting than in the O' perovskite phase belonging to the space group $D_{2h}{}^{16}$-P*bnm* [33]. Recently discussed for $NdMnO_3$ , [20] that environment trails its origin to the orbital disordered primordial perovskite lattice in which background $e_g$ electrons are in fluctuating d-orbitals [34, 35]. The consequence of orbital electron dynamics in hybridized oxygen-manganese distorted $e_g$ orbitals implies embedded minute orbital misalignments. This effect randomizes non-static electrons in the likely breathing distortion of an enhanced correlated, strongly polarizable, metal-oxygen bond creating a THz instability outcome from the spontaneous strain distortion.[36] In $TmMnO_3$ the collective instability, shown in Fig. 7, appears as a room temperature smooth strong tail below 100 $cm^{-1}$.[37,38]

We found that on cooling the profile change as a consequence of gradually localizing electrons inducing long-range magnetic order. The net electric dipole of this collective excitation traced to loosely bond $e_g$ electrons in a d-orbital fluctuating environment it is compromised by small changes in the bipyramidal tilting and the O-Mn-O angle. Magnetically disordered electrons in fluctuating orbitals localize via Coulomb interactions leading to the ambient unstructured broad band. On cooling toward $T_N$ the electrons will exhibit increasing charge and magnetic short-range correlations, and at the same time, orbital fluctuations will slow down due to exchange adding to Coulomb interactions. Oscillating electrical dipoles condense at $T_N$ primarily into two smooth soft infrared active bands that harden as the long range magnetic order sets in. The now magnetically tangled electrons have their motion prevailing over the paramagnetic spin entropy, having this, as the leading factor for inducing gradual magnetic ordering in the bands onset well above nominal $T_N$.

At 4 K, long-range E-type magnetism order is well established in the ***ab*** planes and thus band profiles. It then follows that those excitations, within the macroscopic framework of the generalized LST relation taking into account the coupling of electric and magnetic fields, may be considered direct consequence of magnon coupling magnetoelectrically with individual constants $\alpha_i(\omega)$ (i=1,2) [24] conforming quasiparticles assigned to hybrid-electric dipole (lattice)-spin (Goldstone) modes.[40,41] Their temperature dependence is tuned to the development of long-range magnetic order below $T_N$.

Fig. 7 also shows that band change from above to below $T_N$ is continuous as for a second order displacive phase transition. Second order transitions are characterized by critical exponents. We found these critical exponents by adjusting the experimentally measured peak frequencies to a power law given by

$$\omega_{soft} = A \cdot \left(T_{Cr} - T\right)^{\beta} \tag{9}$$

being A a constant, and $T_{Cr}$ an effective critical temperature.

Peaking at 26 $cm^{-1}$ the band at lower frequency is Rare Earth dependent.[42,37] (Fig. 8) Being sharp and well defined we simply assign peak positions based on the maximum in the spectra. The three parameter power fit to the temperature dependent peaks (Fig, 9), yields $\mathbf{A_{Sp1} = 4.34 \pm ^{0.31}_{0.01}}$, $\mathbf{T_{Cr} = 84 \pm ^{1}_{3}}$ with $\mathbf{\beta_{Sp1} = 0.420 \pm ^{0.001}_{0.001}}$ where a plus sign generates the upper and the minus sign the lower curve of a confidence band in every of the three sets of experimental points shown in figure 9 (Spj stands for the jth soft peak). β ~0.42 matches the critical exponent for spontaneous long-range antiferromagnetic order in a hexagonal lattice. [43] Poirier et al [44] reported on the presence of spin fluctuations above and below the critical temperature of hexagonal $YMnO_3$. Exchange interactions coupled strongly to the lattice yield conventional antiferromagnetic long-range order with β= 0.42. We find this coincident with our $TmMnO_3$ spectra (Fig. 7) and accordingly, as in $NdMnO_3$, we call that excitation *spin-like.*

The second asymmetric soft band centered at ~50 $cm^{-1}$ (Fig. 7) is similar to the one discussed for perovskite $NdMnO_3$. [37]. However, as it is shown in Fig. 8 (a) the band profile for that excitation appears as an envelop of the corresponding in $TmMnO_3$ in which, at difference of $NdMnO_3$, there is a band splitting suggesting a more complex picture. Fig. 8(a) shows that weakest component of the two at ~54 $cm^{-1}$ and 4 K matches the zone center magnon found in E-type $LuMnO_3$ [45] at the same energies, while the second one peaking at ~36 $cm^{-1}$, although being the dominant feature of the pair, does not have a direct correlation with the magnetic dispersion.

On the other hand, Petit el al have found in hexagonal $YMnO_3$ a gap that opens below $T_N$ in the transverse acoustic phonon branch at $q_0$~0.185. The phonon dispersion is polarized along the ferroelectric axis **c**.[17] The magnon-phonon coupling creates hybridized quasiparticles revealing a common channel for antiferromagnetism and polar order. We found, that the infrared band splitting of $TmMnO_3$ at ~50 $cm^{-1}$ (~6 meV) agrees with the hybridized mode magnon-acoustic phonon anomaly. As it is shown in Fig. 8 (b) the reflectivity of $TmMnO_3$ at 12 K reproduces remarkably well the neutron intensity profile between 30 $cm^{-1}$ and 80 $cm^{-1}$.

That is, our findings matching infrared active and magnons suggest, as it is shown in figure 8(a), that the two zone center neutron detected magnons have a direct one to one counterpart with two, out of three, bands in the THz region. The fact that these three bands are detected in reflectivity in the far infrared (figure 8 (a)) as zone center features is reminiscent to experimental partial detection of a phonon density states optically active due to defect impurities introduced in the lattice in minute quantities. Intrinsic orbital fluctuations would play this role in $TmMnO_3$. The sharp band shape at ~26 $cm^{-1}$ would then be a consequence of a zone center flat ending dispersion The split envelope of the band at ~50 $cm^{-1}$ resulting in two much broader bands would be a consequence of the near zone center strong curvature of the corresponding dispersion combined with the fluctuating environment outcome of the acoustic mode gap opening at $T_N$.

This contrasts with an electromagnon interpretation, i.e. magnons excited by the ac electric field component of the light, on the origin of what seems to be the same bands. This view is shared by a number of earlier publications reporting on low frequency infrared absorption resonances in hexagonal $RMnO_3$ (R=Rare Earth) (see discussion in [37] and refs. there in). Our measurements allow to infer that understanding electromagnons as unique transverse features induced by infrared radiation ought to be reexamined since the THz band activity can be inferred to be consequence of the extraordinary magnetoelectric couplings in a highly anharmonic environment result of orbital fluctuations .

Fig. 9 shows power law fits for the two soft bands centered at ~50 $cm^{-1}$ We use asymmetric Weibull shapes for individualizing band peak frequencies.[37] These yield $\mathbf{A_{Sp2} = 11.80 \pm {}^{0.00}_{0.34}}$ , $\mathbf{T_{Cr} = 84. \pm {}^{2}_{4}}$ with $\mathbf{\beta_{Sp2} = 0.250 \pm {}^{0.008}_{0.002}}$ and $\mathbf{A_{Sp3} = 23.00 \pm {}^{0.32}_{0.30}}$ , $\mathbf{T_{Cr} = 84. \pm {}^{1}_{0}}$ with $\mathbf{\beta_{Sp3} \approx 0.20 \pm {}^{0.000}_{0.001}}$ for the split main and weaker band, respectively. β=0.25 is the exponent for critical regimes close to temperatures at which a structural phase transition takes place. It suggests a delicate balance between a discontinuous first order and a continuous second order phase transition, i.e. where the fourth power coefficient, in the free energy expanded in terms of the order parameter, is zero. This is the line where the first order transition meets the line of the second order transition.[46,47] The β=0.25 that holds in a fairly large temperature range below $T_N$ is indicative that $TmMnO_3$ is indeed affected by a structural fluctuation.

This fact also allows a more comprehensive understanding of those excitations as it becomes more transparent from their critical association with the transverse acoustical phonon ferroelectric-magneto coupling thus justifying a "phonon-like" name. In this fashion one is able to trace, as in $YMnO_3$,[17] through the THz peak temperature dependence (Fig. 9) the opening of the transverse acoustical phonon gap at about $T_N$ showing that the perturbation introduced at finite **q** by the strong coupling between phonon and spin waves results for $TmMnO_3$ in distinct zone center hybrids. All modes obey power laws with $T_N$ as the critical temperature.

Our findings also shade a better understanding of the same but stronger feature found in orthorhombic $NdMnO_3$ (Fig. 8 (a)) where, although not having substructure, it is also associated with zone center magnons.[37] The TA gap opening is symmetry dependent on spin and phonon dispersions and only when there is band crossing as in $YMnO_3$[17] the strong magnetoelastic coupling materializes the multiferroic character. In hexagonal $TmMnO_3$, it is inferred as zone center infrared active features. And for this very reason, in orthorhombic $NdMnO_3$ those two hybridised excitations remain undefined in spite of having infrared activity that it is favored by the underlying reduced correlation lengths in the orbital fluctuating spin frustrated lattice. This does not prevent, however, that both compounds, $NdMnO_3$ and $TmMnO_3$, share at low temperatures band hardening tuned to the developing of long-range magnetic order in the Néel phase.

Our results presents another view of a second order transition in strongly correlated oxides for which the long-range macroscopic view given by X-ray patterns, and described by a lattice space group, is not required to change.

## CONCLUSIONS

Summarizing, we discussed the temperature dependent far infrared 1-E and reflectivity spectra of $TmMnO_3$ from 1910 K to 4 K.

At the highest temperature we found that the number of infrared bands are lower than that the predicted for centrosymmetric $P6_3/mmc$ ($D_{6h}^4$) (Z=2) lattice. We reason that this might be due high temperature anharmonicity band broadening without excluding the possibility of defect dynamic contribution induced by bipyramidal misalignments. On cooling, at 1600 K ± 40 K, $TmMnO_3$ evolves from non-polar to antiferroelectric-ferroelectric polar phase in a two step lattice transition. The ferroelectric distortion in the 800 K to 600 K range is apparent as phonon bands where Mn-O-Mn motions become more intense.

Room temperature reflectivity is fitted using 19 phonons and this number is maintained down to 4 K.

A low energy collective excitation is identified as a THz instability found at room temperature associated with eg electrons in a d-orbital fluctuating environment. It condenses into two modes that correlate to known zone center magnons. One, peaking at 26 $cm^{-1}$ is dependent on the Rare Earth and long-range antiferromagmetic order in the hexagonal lattice while the higher frequency companion at ~50 $cm^{-1}$ splits into two bands. The weaker band at 54 $cm^{-1}$ can be assimilated to the upper branch of the gap opening in the transverse acoustical mode polarized along the ferroelectric axis found in $YMnO_3$ [17]. The stronger second component centered at 36 $cm^{-1}$, with $\beta$=0.25 as for critical regimes close to structural phase transitions, corresponds to the phonon-like lower branch of the same TA gap, creating truly zone center magnetoelectric hybrid quasiparticles of mixed character. Their temperature dependence also allows concluding that the gap opening takes place at ~$T_N$ confirming the ferroelectric-magnetic coupling in the magnetic ordered phase. Overall, our measurements give a comprehensive view of the evolution of $e_g$ electrons entangled d-orbitals giving rise to a low energy collective instability and ferroelectric-magnetic hybridized zone center modes. This scenario thus allows us to associate observations of electronically induced mechanisms for colossal magnetoresistance or polar ordering in transition metal oxides involving orbital/charge and/or spin fluctuations. [48, 49] We conclude that in perovskite $NdMnO_3$ the equivalent picture represents a continuous instability which might be driven by an external field to transform $NdMnO_3$ into a multiferroic compound by perturbation enhancing the TA phonon-magnetic correlation.

We also hope that our measurements will motivate the need for inelastic neutron scattering measurements in order to help to quantify the individual mechanism driving multiferroic oxides.

## ACKNOWLEDGEMENTS

NEM is grateful to the CNRS-C.E.M.H.T.I. laboratory and staff in Orléans, France, for research and financial support in performing far infrared measurements. LdelC and NEM thanks the Berliner Elektronenspeicherring -Gesellschaft für Synchrotronstrahlung–BESSYII- for financial assistant and beamtime allocation under project# 2013-1-120813. NEM also acknowledges partial financial support (PIP 0010) from the Argentinean Research Council (Consejo Nacional de Investigaciones Científicas y Técnicas-CONICET).. Funding through Spain Ministry of Economy and Competitivity (Ministerio de Economia y Competividad) under Project MAT2013-41099-R is acknowledged by JAA and MJML

# Table 1

Dielectric simulation fitting parameters for $TmMnO_3$

| T (K) | $\varepsilon_\infty$ | $\omega_{TO}$ (cm$^{-1}$) | $\Gamma_{TO}$ (cm$^{-1}$) | $\omega_{LO}$ (cm$^{-1}$) | $\Gamma_{LO}$ (cm$^{-1}$) |
|---|---|---|---|---|---|
| | | 195.7 | 151.1 | 237.7 | 114.8 |
| | | 354.8 | 137.2 | 418.4 | 243.5 |
| 1760 K | 2.25 | 428.3 | 273.0 | 532.8 | 195.5 |
| | | 667.2 | 179.2 | 682.5 | 135.02 |
| | | 1681.6 | 2926.5 | 2609.4 | 7112.9 |
| | | | | $\omega_{pl}$ | |
| | | | 5712.8 | 771.0 | 3456.9 |

| T (K) | $\varepsilon_\infty$ | $\omega_{TO}$ (cm$^{-1}$) | $\Gamma_{TO}$ (cm$^{-1}$) | $\omega_{LO}$ (cm$^{-1}$) | $\Gamma_{LO}$ (cm$^{-1}$) |
|---|---|---|---|---|---|
| | | 89.0 | 104.2 | 93.7 | 126.4 |
| | | 170.8 | 83.4 | 176.5 | 149.6 |
| | | 268.1 | 603.9 | 283.7 | 399.7 |
| 1382 | 2.41 | 377.5 | 81.1. | 381.1 | 42.4 |
| | | 382.7 | 53.2 | 403.5 | 122.0 |
| | | 466.7 | 232.2 | 545.1 | 172.4 |
| | | 650.1 | 128.1 | 688.8 | 125.2 |
| | | 6327.7 | 4276.2 | 6526.0 | 7891.3 |

| T (K) | $\varepsilon_\infty$ | $\omega_{TO}$ (cm$^{-1}$) | $\Gamma_{TO}$ (cm$^{-1}$) | $\omega_{LO}$ (cm$^{-1}$) | $\Gamma_{LO}$ (cm$^{-1}$) |
|---|---|---|---|---|---|
| | | 110.3 | 205.5 | 117.8 | 35.8 |
| | | 217.6 | 41.23 | 221.9 | 46.8 |
| | | 245.5 | 52.5 | 249.1 | 37.8 |
| | | 260.8 | 35.5 | 268.6 | 26.9 |
| 580 | 2.45 | 282.5 | 34.9 | 289.1 | 41.5 |
| | | 302.8 | 120.4 | 345.9 | 70.4 |
| | | 360.1 | 40.2 | 382.4 | 117.8 |
| | | 423.4 | 97.7 | 446.1 | 118.3 |
| | | 485.5 | 294.4 | 521.8 | 37.1 |
| | | 590.4 | 63.4 | 594.5 | 135.2 |
| | | 677.0 | 79.8 | 705.6 | 101.0 |

| T (K) | $\varepsilon_\infty$ | $\omega_{TO}$ (cm$^{-1}$) | $\Gamma_{TO}$ (cm$^{-1}$) | $\omega_{LO}$ (cm$^{-1)}$ | $\Gamma_{LO}$ (cm$^{-1}$) |
|---|---|---|---|---|---|
| 4 | 2.31 | 73.5 | 19.8 | 76.4 | 14.0 |
| | | 93.4 | 53.5 | 103.8 | 62.1 |
| | | 119.4 | 1.5 | 119.6 | 1.5 |
| | | 165.3 | 5.8 | 165.6 | 5.3 |
| | | 220.8 | 17.2 | 222.6 | 44.2 |
| | | 232.7 | 101.4 | 237.4 | 25.4 |
| | | 254.3 | 15.6 | 266.3 | 21.9 |
| | | 279.7 | 18..0 | 281.2 | 7.0 |
| | | 292.0 | 7.8 | 298.4 | 16..5 |
| | | 303.8 | 83.4 | 341.3 | 20.9 |
| | | 358.1 | 22.9 | 368.4 | 16.8 |
| | | 369.7 | 8.1 | 380.7 | 46.7 |
| | | 388.9 | 51.8 | 393.0 | 26.0 |
| | | 395.1 | 28.9 | 420.2 | 10.3 |
| | | 422.3 | 8.0 | 442.8 | 88.0 |
| | | 490.6 | 131.3 | 497.6 | 40.7 |
| | | 518.4 | 69.1 | 524.4 | 19.7 |
| | | 576.7 | 29.6 | 592.1 | 14.5 |
| | | 701.8 | 50.4 | 728.3 | 78.4 |

# FIGURE CAPTIONS

**Figure 1** (color online) (a) Single phase room temperature X-ray (CuKα) diffraction pattern for $TmMnO_3$ fitted in the hexagonal structure, (b) $TmMnO_3$ crystal structure at room temperature defined in the hexagonal space group $P6_3cm$ with the lattice parameters a= 6.0950(1), c= 11.3806(2) Å.

**Figure 2** (color online) $TmMnO_3$ near normal 1-E from 1900 K to 600 K. The mid-infrared spectra below 900 K have been cut because the sample is only partiallly absorbent and transparency induces high levels of noise. Dashed arrow points to the highest frequency phonon temperature evolution into the structural phase transition. For better viewing, the spectra have been displaced vertically by 0.10 relative to each other. Inset: phonon profile changes at the centrosymmetric non-polar to hexagonal polar antiferroelectric-ferroelectric phase transition at 1600 K ± 40 K.

**Figure 3** (color online) $TmMnO_3$ 1-E and reflectivity spectra (table 1) plotted in semilog scale (circle: experimental, full line: fit) at (a) 1760 K, (the oxidation induced extrinsic carrier contribution is shown as an overdamped Drude contribution in dashed line); (b) 1382 K ; (c) 580 K ; (d) 4 K.

**Figure 4** (color online) Near normal reflectivity and 1-E in the phonon region. Arrows point to Mn-O-Mn vibrational groups denoting the gradual lattice change associated with the inversion center loss rendering the ferroelectricity onset below ~700 K. For better viewing the spectra have been displaced vertically by 0.10 relative to each other.

**Figure 5** (color online) Far infrared temperature dependent reflectivity of $TmMnO_3$ from ambient to 4 K. For better viewing, the spectra have been displaced vertically by 0.10 relative to each other.

**Figure 6** (color online) Upper panel: temperature dependent band profile of the 300 K ~210 $cm^{-1}$ lattice vibrational mode. (circle: experimental points, full line: fit). We found that a better fit at ~$T_N$=84 K requires an extra side band oscillator. Lower panel: relative oscillator strength (eq (6)) of the same mode and the extra one introduced in the Néel phase (see text)

**Figure 7** (color online) Lowest frequency instability due to d-orbital $e_g$ fluctuations from 300 K to 4 K. Note that the onset of band condensation, into two hybrid bands and mode hardening starts well above ~84 K meaning that magnetic correlations exists above nominal ~$T_{N.}$ For a better viewing, the spectra have been vertically shifted by 0.10 relative to each other.

**Figure 8** (color online) (a) Right panel: particular of the magnetic dispersion of $LuMnO_3$ reported by Lewtas et al [45] Left panel: Soft bands at 4 K for $TmMnO_3$ (square) and $NdMnO_3$ (triangle). Note that the $NdMnO_3$ asymmetric higher frequency band appears as an envelope of the $TmMnO_3$ split band. It suggests an existing perturbation but lacks, due to the orthorhombic topology, the band splitting induced by the transverse acoustic phonon gap opening and coupling of the hexagonal counterpart found for $YMnO_3$ by Petit et al [17], (b) zone center THz $TmMnO_3$ hybridized split band at 12 K (square) as it compares to the neutron profile (dot-bar) for the same excitation at q=0.15 in $YMnO_3$ [17]. Dashed lines are a two gaussian fit to this data by [17].

**Figure 9**. (color online) Power law fits for the phonon-like and spin-like modes (see text). Dashed lines: Temperature dependent fit extrapolation showing that dispersion crossing, and thus gap opening, may be traced up to ~$T_{N.}$

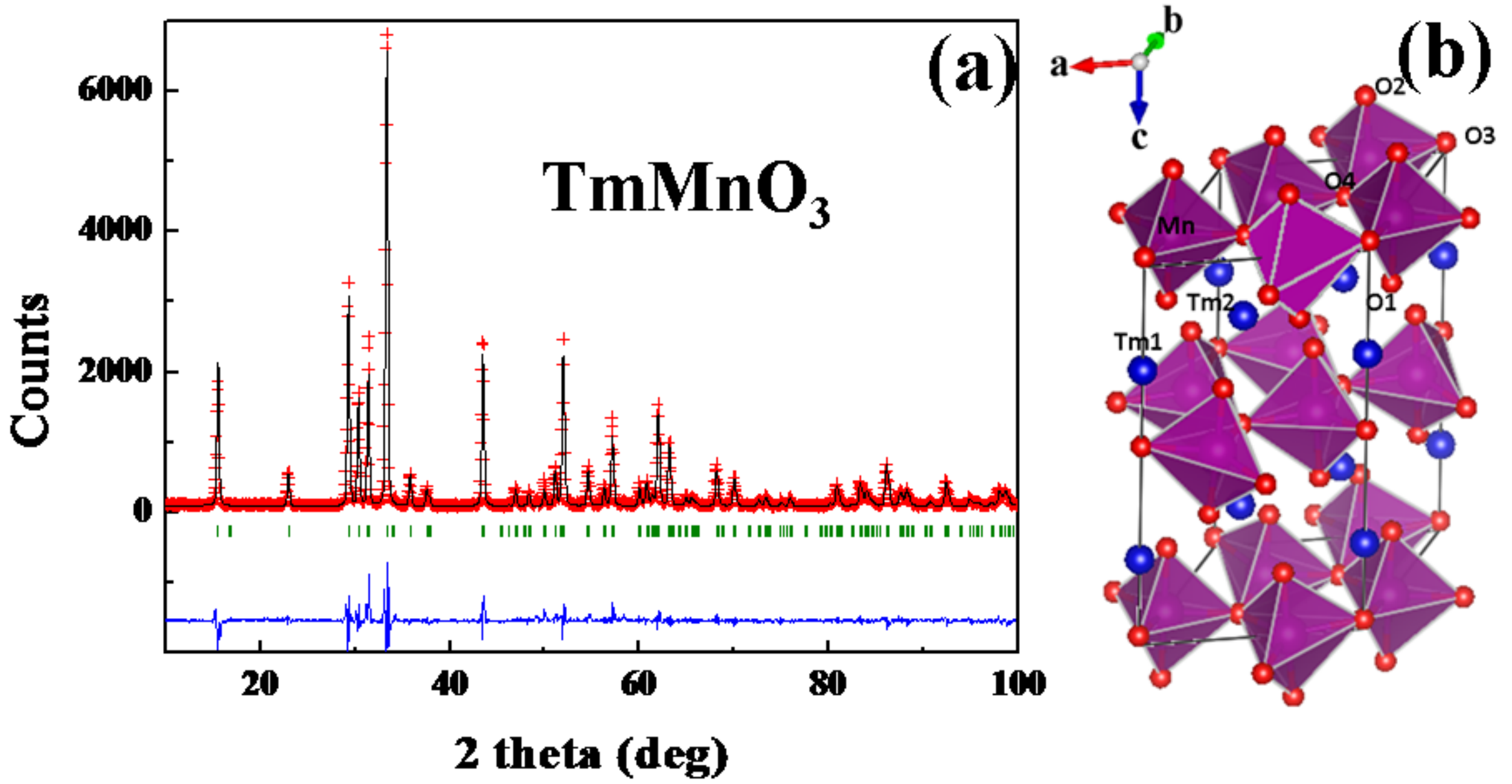

# Figure 1

**Massa et al**

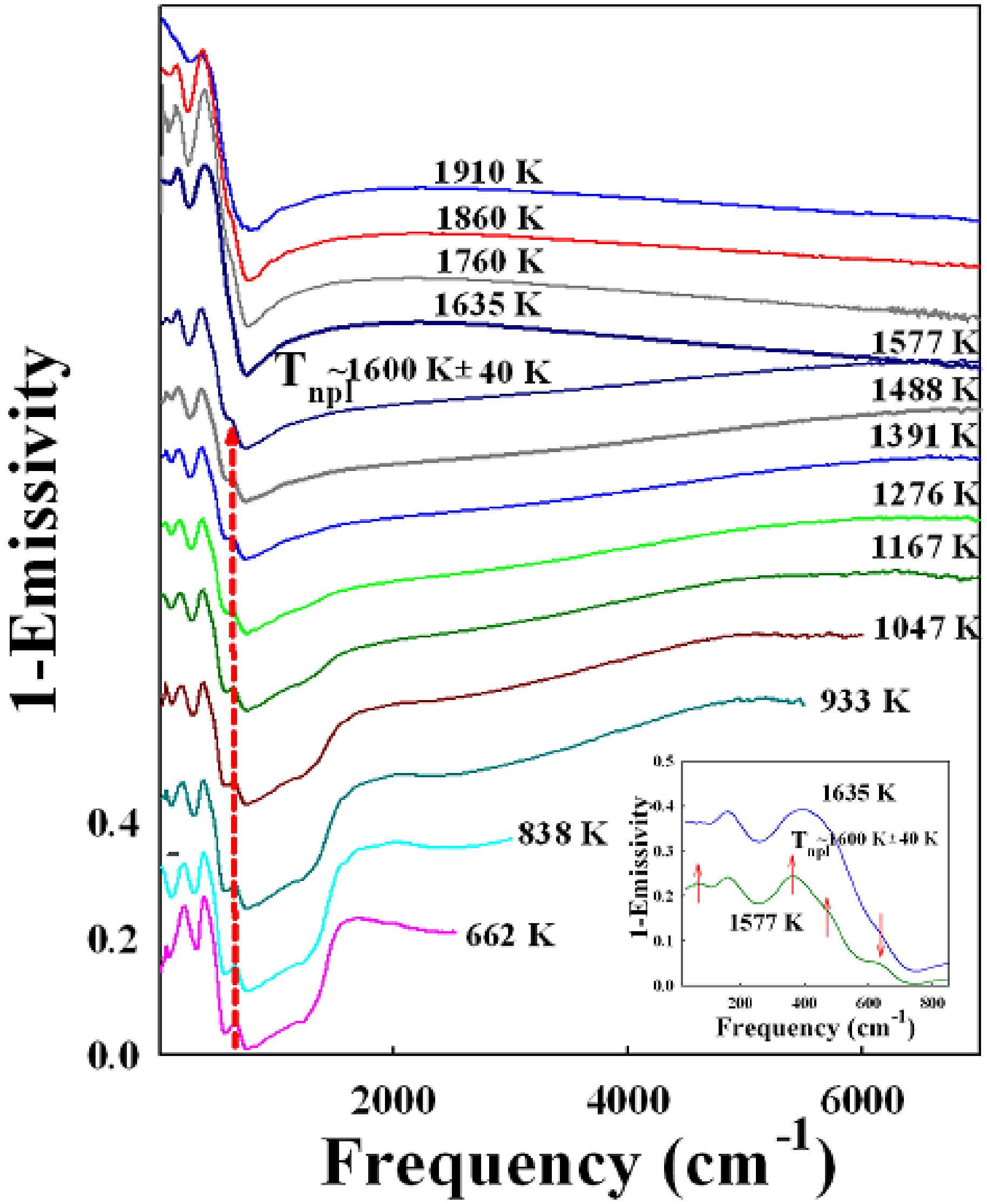

# Figure 2

# Massa et al

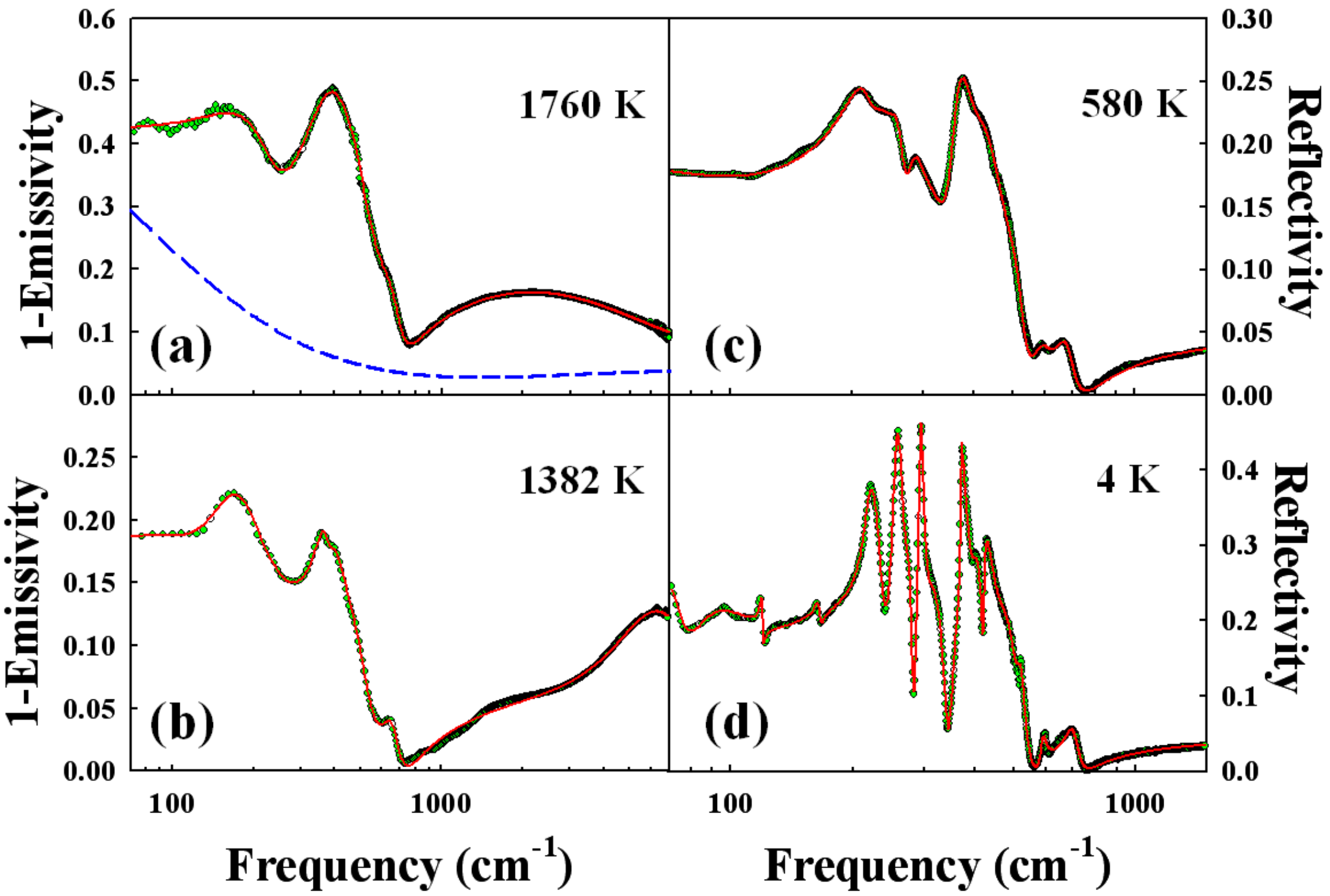

# Figure 3

## Massa et al

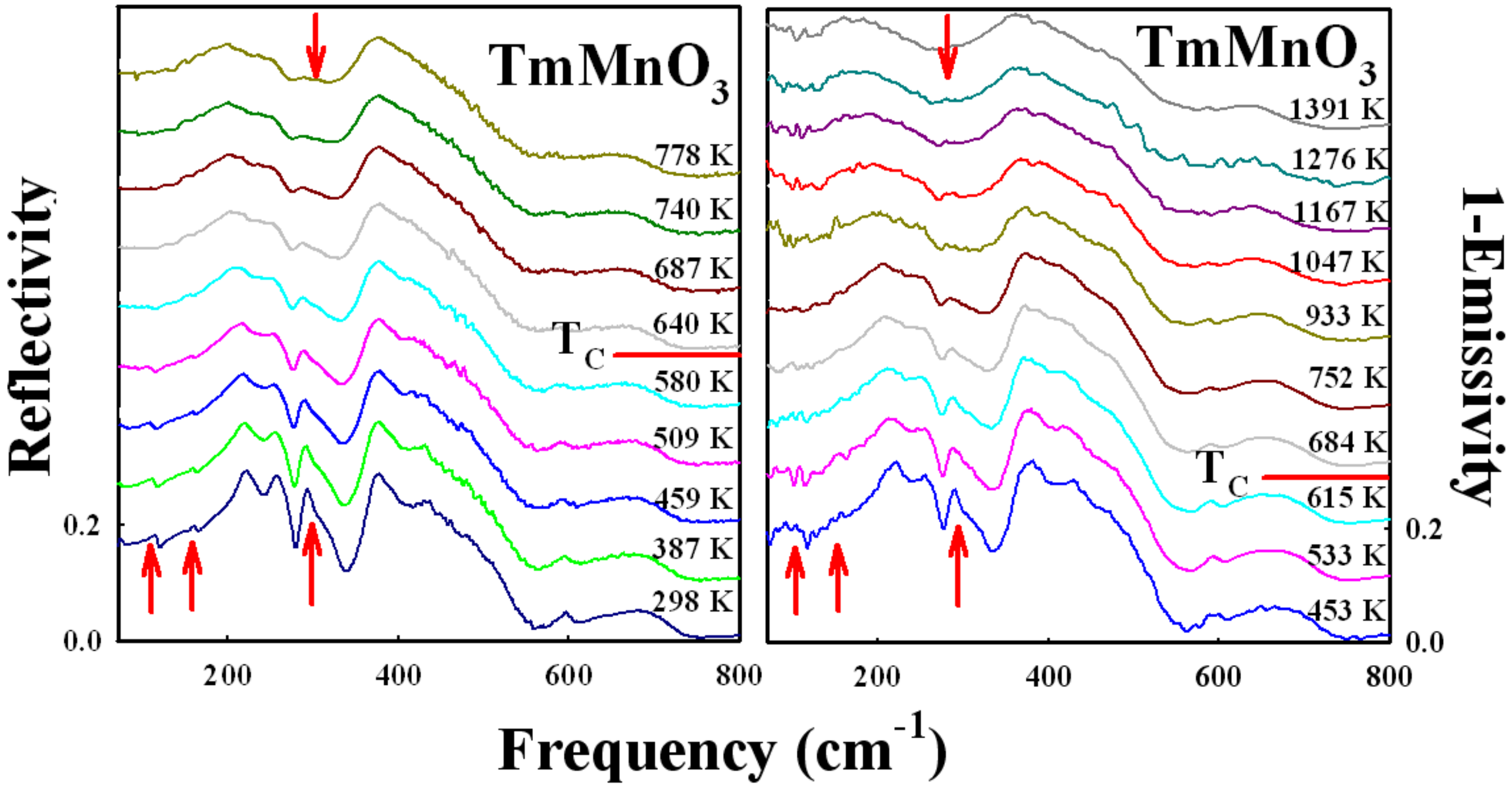

# Figure 4

# Massa et al

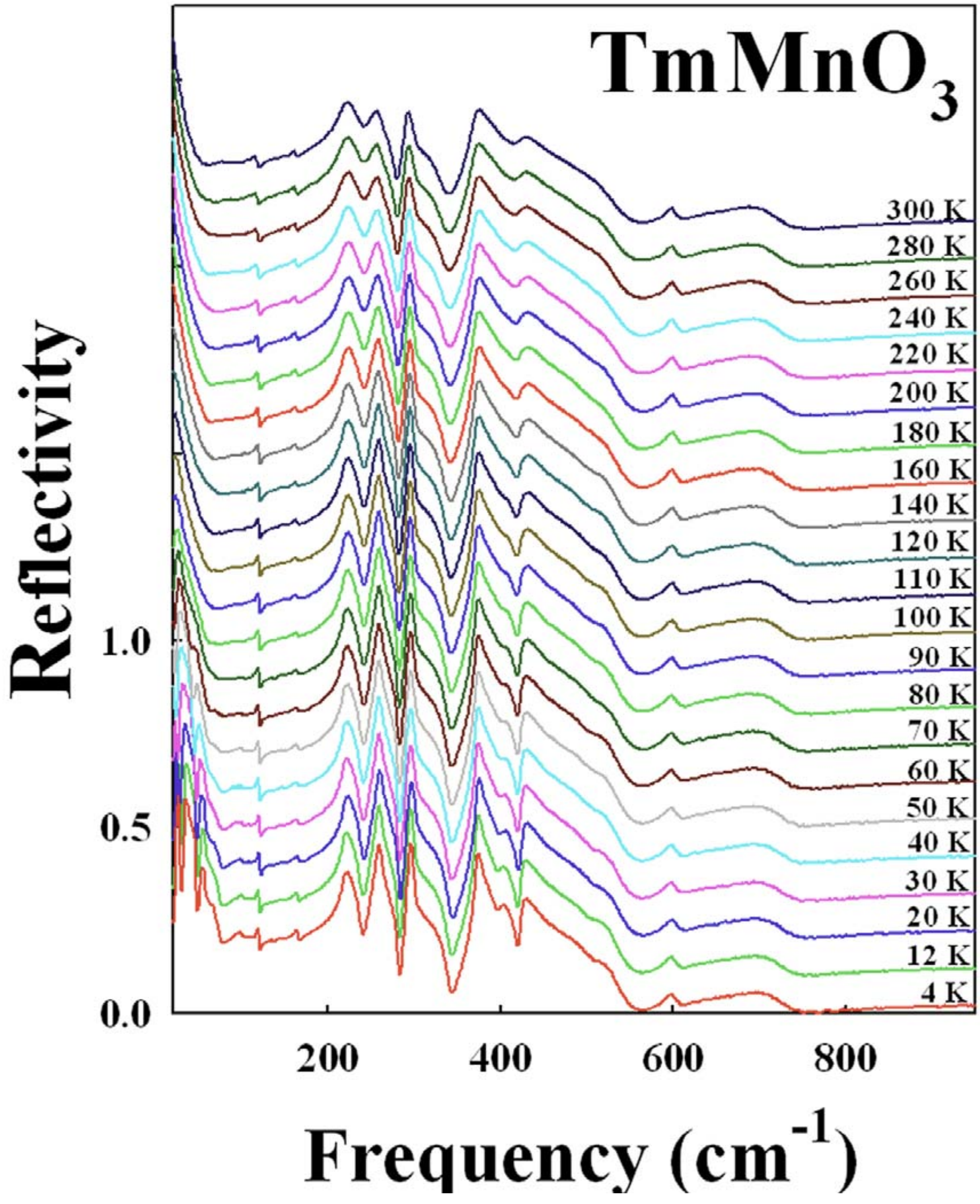

**Figure 5**

**Massa et al**

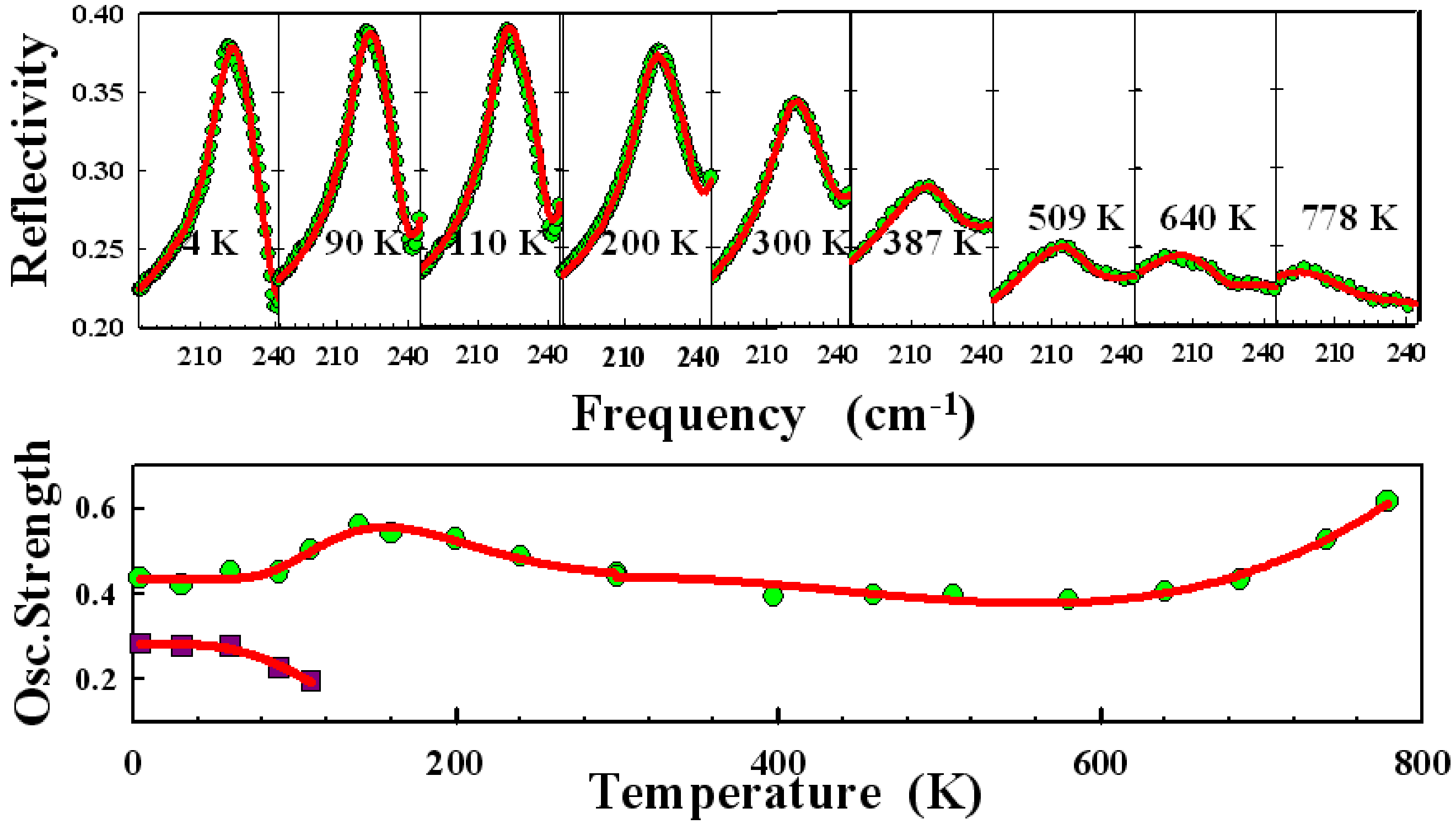

**Figure 6**

**Massa et al**

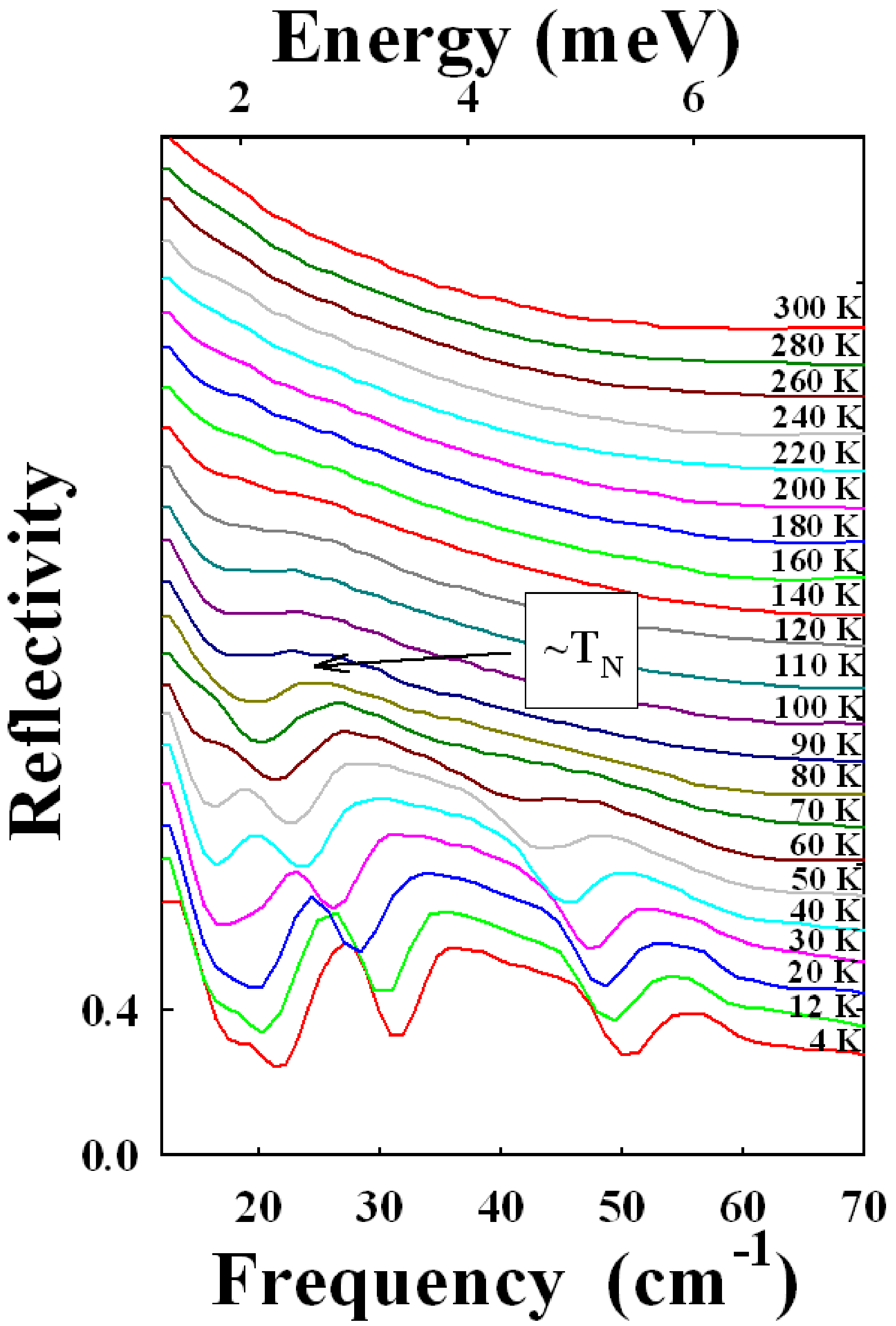

**Figure 7**

**Massa et al**

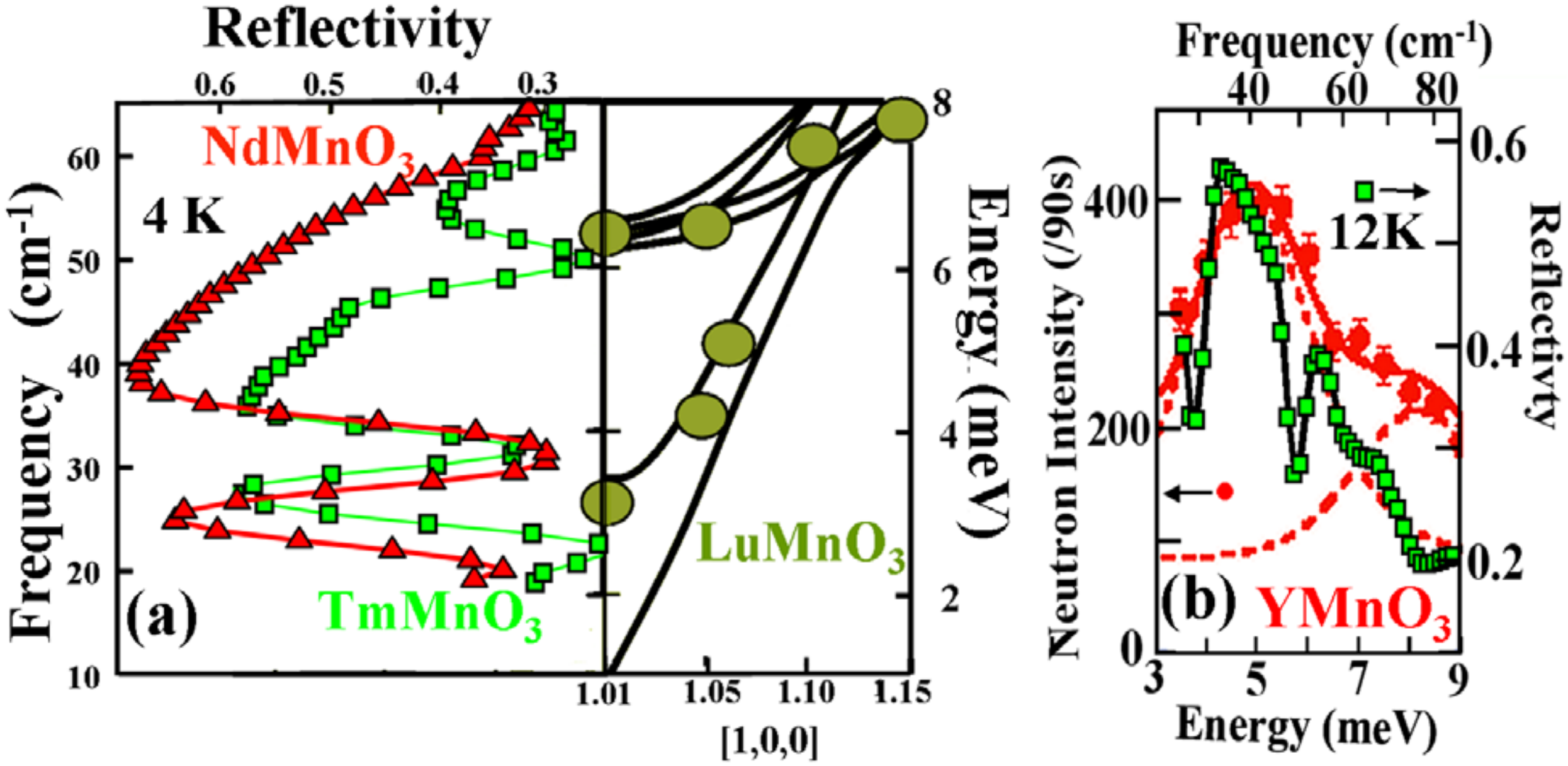

# Figure 8

## Massa et al

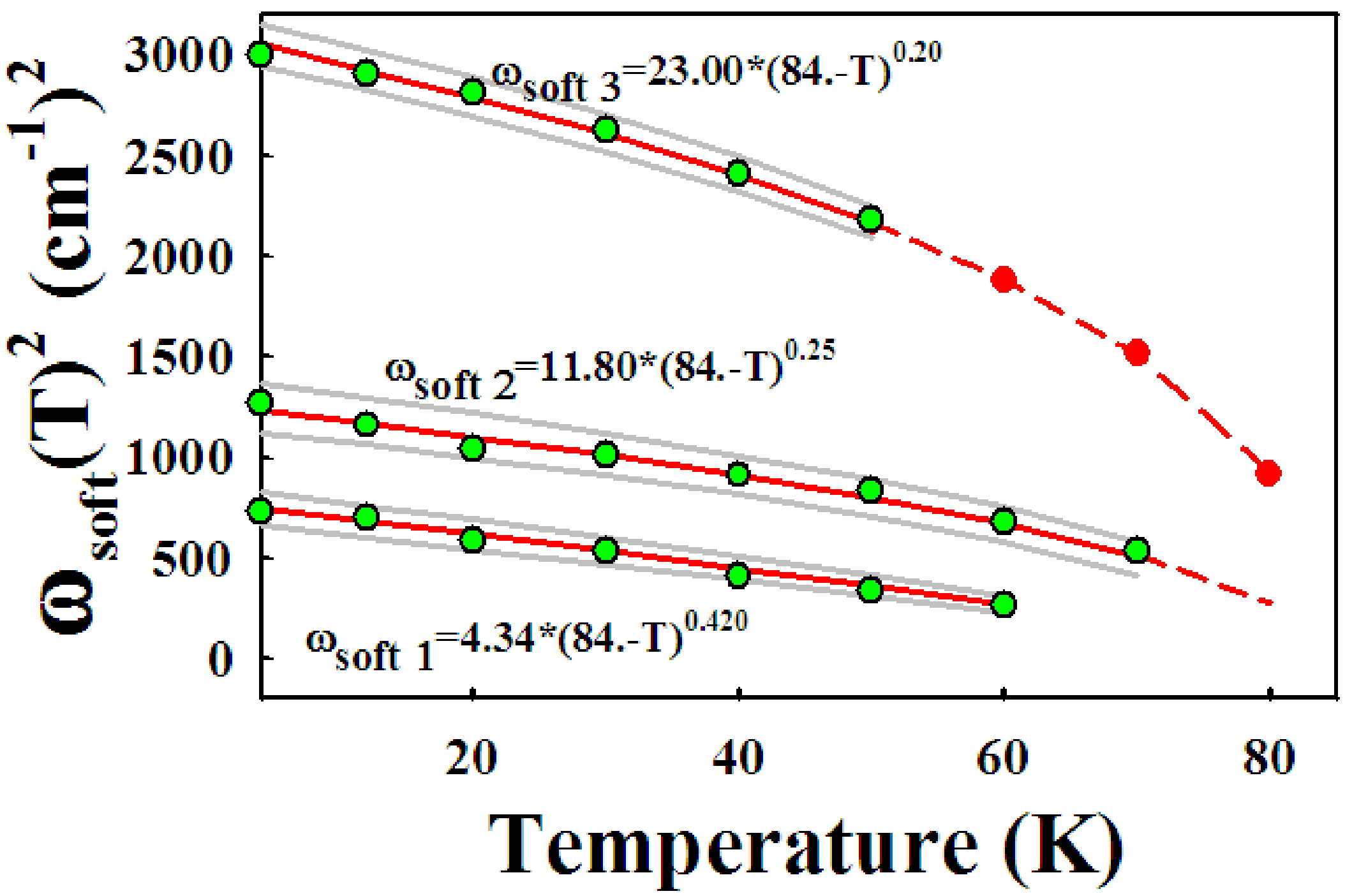

# Figure 9

# Massa et al